\documentclass[%
aps,
pre,
reprint,
amsmath,
amssymb,
superscriptaddress]{revtex4-2}
\usepackage{graphicx}
\usepackage{xcolor}
\usepackage{bm}
\usepackage{siunitx}
\usepackage[%
hidelinks,
colorlinks=true,
citecolor=blue,
urlcolor=blue,
linkcolor=blue
]{hyperref}

\definecolor{red}{HTML}{C74431}

\definecolor{green}{HTML}{3C9455}

\begin{document}
\author{Roland Wiese}
\email{wiese@itp.uni-leipzig.de}
\author{Klaus Kroy}
\affiliation{Institute for Theoretical Physics, Leipzig University, 04103 Leipzig, Germany}
\author{Viktor Holubec}
\email{viktor.holubec@mff.cuni.cz}
\affiliation{Faculty of Mathematics and Physics, Charles University, CZ-180 00 Prague, Czech Republic}
\date{\today}

\title{Numerical bacterial heat engine: Average performance and effective temperature}
\title{Average performance and effective temperature of \emph{in silico} bacterial heat engines}
\title{Average performance and effective temperature of active heat engines}
\title{Modeling the Efficiency and Effective Temperature of Bacterial Heat Engines}
\begin{abstract}
We present a Brownian dynamics simulation of the bacterial Stirling engine studied by \emph{Krishnamurthy et al., Nat. Phys. {\bf 12}, 1134 (2016)}.
In their experimental setup, an overdamped colloid in an optical trap with time-modulated stiffness interacts with a bacterial bath that we represent by an ensemble of overdamped active particles.
In the parameter regime of the experiment the thermodynamic performance is governed by an effective temperature and can be parametrized analytically by active Brownian particle models.
We quantitatively reproduce the efficiencies reported for the experiments under the assumption that energy exchange with the bath during isochores of the cycle is entirely recuperated and in agreement with the second law.
\end{abstract}
\maketitle
\section{Introduction}
Thermodynamic heat engines convert disordered forms of energy or ``heat'' into systematic work.
Their efficiency is limited by universal temperature-dependent bounds dictated by the second law of thermodynamics.
These extend to non-classical baths, be they quantum-mechanical~\cite{Rosnagel2014, Scully2003}, relativistic or ``active''~\cite{krishnamurthy2016micrometre,holubec2020active-brownian-heat-engines,holubec2020underdamped}, as long as their energy supply can be characterized by an effective temperature, which seems a \emph{sine qua non} condition for speaking of a heat bath, at all. 
\begin{figure}[t]
\centering
\includegraphics[width=\linewidth]{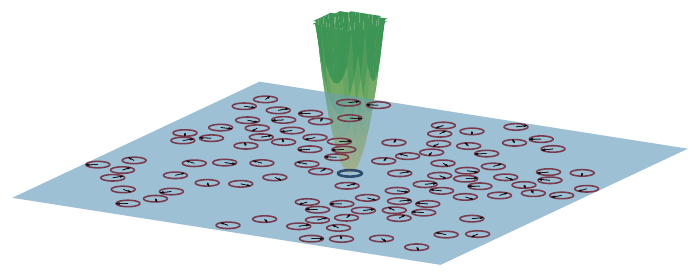}
\caption{Snapshot of the simulated two-dimensional setup: 
A passive Brownian probe (blue disk) confined by a harmonic potential (green cone) diffuses in an equilibrium solvent (blue, modeled as a standard Markovian heat bath) containing active Brownian particles (red circles) to represent the active bacterial bath component employed in the experimental setup of Ref.~\cite{krishnamurthy2016micrometre}. 
}
\label{fig:setup}
\end{figure}

Recent developments in micro-manipulation techniques now allow the realization of heat engines based on a single Brownian particle \cite{martinez2017colloidal-heat-engines-review}. 
A remarkable engine of this kind~\cite{krishnamurthy2016micrometre} is based on a particle immersed in a solution of living bacteria, serving as a paradigmatic example of a non-equilibrium (active) bath. 
The authors of Ref.~\cite{krishnamurthy2016micrometre} reported that they operate their engine along a Stirling cycle, with efficiencies surpassing the corresponding ultimate thermodynamic bound attained for diverging bath temperature. 
This result is perplexing because the gain in efficiency seemingly cannot be explained just by the large effective temperature reached by virtue of the broken zeroth law (due to the bacterial activity that does not thermalize with the surroundings)~\cite{holubec2020active-brownian-heat-engines}.

The authors of the experimental paper argued that their results follow from the non-Gaussianity of the noise~\cite{krishnamurthy2016micrometre}, but this explanation was later shown to be insufficient~\cite{holubec2020active-brownian-heat-engines, park2022effects-of-non-markovianity}.
During the past few years, several studies attempted to crack the puzzle either by considering the dependence of the effective temperature on stiffness~\cite{holubec2020active-brownian-heat-engines}, a complicated memory of the active bath~\cite{zakine2017stochastic-stirling,park2020active-reservoirs,park2022effects-of-non-markovianity, chang2023stochastic-heat-engines, albay2023engineered-active-noise}, or the temperature dependence of the bacterial activity~\cite{Kwon2024}. 
But none of these explanations seems entirely convincing, as very specific dependencies of the effective temperature or non-Markovian effects are required to reproduce at least a part of the observed efficiency as a function of the effective temperature~\cite{krishnamurthy2016micrometre}.

To resolve the issue, we performed computer simulations faithful to the experiments described in Ref.~\cite{krishnamurthy2016micrometre}. 
An overdamped colloidal probe particle is trapped in a harmonic potential and diffusing in a solution of generic active Brownian particles, as sketched in Fig.~\ref{fig:setup}. 
We tuned the model parameters according to the experiments, including the Stirling-like time-dependent protocols for trap stiffness, bath temperature, and activity. 
As in the experiments, the resulting cycles were quasi-static so that the effective temperature $T_{\rm eff} = k\sigma / 2k_B$ is simply given by the product of the position variance $\sigma$ and the trap stiffness $k$~\cite{holubec2020active-brownian-heat-engines, puglisi2021optimization}.
We checked numerically that $T_{\rm eff}$ can be obtained with reasonable precision when sampling the data with the measurement frequency applied in the experiments (our first main result, shown in Fig.~\ref{fig:var-t-and-Teff-k}). 
The resulting $T_{\rm eff}$ depends on the trap stiffness $k$ so that the actual cycle in the $T_{\rm eff}-k$ plane is no longer of the Stirling type. 

The efficiency $\eta = W_{\rm out}/Q_{\rm in}$ of the engine is defined as the output work per cycle, $W_{\rm out}$, over the total heat input $Q_{\rm{in}}$, i.e., the energy accepted from the active bath during the isochoric expansion and the hot isotherm.
It falls below the upper limit of the Stirling efficiency, $1/[1~+~1/\log(k_{\max}/k_{\min})]$, reached for an infinite temperature difference between the hot and cold reservoirs (where $k_{\rm max}$ and $k_{\rm min}$ are the upper and lower bounds on the trap stiffness during the cycle). 
It is thus significantly smaller than the efficiency reported in Ref.~\cite{krishnamurthy2016micrometre}. 
On the other hand, when we calculate the efficiency as $\eta^\star = W_{\rm out}/Q_{\rm in}^\star$, where $Q_{\rm in}^\star$ is merely the energy accepted from the active bath during the hot isotherm alone, we obtain perfect agreement with the experimental data, shown in Fig.~\ref{fig:eta-vs-Teff}, as our second main result. 
This strongly suggests that the authors of Ref.~\cite{krishnamurthy2016micrometre} applied this alternative definition of efficiency.
It pertains to Stirling cycles with a perfect recuperation mechanism of the heat interchanged during the isochoric branches and is only bounded by Carnot's efficiency for reversible cycles. 
While the authors of Ref.~\cite{krishnamurthy2016micrometre} indeed state in the methods section that they define efficiency as $\eta^\star$, their whole analysis assumes that it should be bounded by the Stirling efficiency, which actually requires the alternative definition $\eta$.
That many readers apparently understood that the actual efficiency definition employed in Ref.~\cite{krishnamurthy2016micrometre} was $\eta$ could also stem from the fact that, while $\eta^*$ is often a reasonable definition for conventional, macroscopic Stirling engines, it is difficult to conceive possible recuperation mechanisms for the energy interchanged with bacterial baths, rendering $\eta^{\star}$ an inappropriate figure of merit, in this case. \textcolor{black}{However, we note that a first step toward energy recuperation from an active bath was made in a recent experimental study~\cite{Ginot2024}, where the authors successfully recovered energy from a viscoelastic bath using a moving optical potential. It remains to be seen whether these results can be generalized to baths composed of active Brownian particles and to protocols with a breathing harmonic potential, as used in the bacterial heat engine.}

We found that the predictions for reversible efficiencies of the Carnot and Stirling cycles, expressed in terms of the (correctly defined) effective temperature difference $\Delta T_{\rm eff}$, are consistent with the efficiencies obtained by directly measuring input energy and output work both in simulation and experiment, contrary to the conclusion in Ref.~\cite{krishnamurthy2016micrometre}.
Moreover, as we show, the measured profiles of the effective temperatures during the cycle can very well be fitted, both for our simulations and for the experiments~\cite{krishnamurthy2016micrometre}, by the effective temperature of either an active Brownian particle or an active Ohrstein-Uhlenbeck particle, trapped in a harmonic potential.
Both toy models can thus serve as effective models for the passive tracer together with its active bath. 
This allows us to obtain analytical formulas for $\eta$ and $\eta^\star$, shown to nicely agree with our numerical data and the experimental data from Ref.~\cite{krishnamurthy2016micrometre}, in Fig.~\ref{fig:eta-vs-Teff} below.

The independent question, which of these simple effective toy models can better reproduce other features of the setup beyond its average thermodynamics, chiefly its stochastic thermodynamics, is addressed in an accompanying paper~\cite{partII}.

The rest of the paper is structured as follows.
In the subsequent Sec.~\ref{sec:model}, we detail our numerical model, the protocol, and the experimentally motivated parameter values. 
Section~\ref{sec:theory} contains definitions of work, heat, efficiency, and effective temperature of the studied setup.
Furthermore, it contains analytical results for these variables obtained using the mentioned effective toy models, namely the active Brownian particle model and the active Ohrstein-Uhlenbeck model.
Readers familiar with these theoretical aspects of the analysis of active Brownian heat engines can skip the theoretical part and directly consult our simulation results in Sec.~\ref{sec:results}. We conclude in Sec.~\ref{sec:conclusion}.

\section{Model of a bacterial heat engine}
\label{sec:model}
The experiment described in Ref.~\cite{krishnamurthy2016micrometre} utilizes a quasi-two-dimensional setup, where optical tweezers are employed to trap a colloidal probe that diffuses in an aqueous solution of bacteria, as illustrated in Fig.~\ref{fig:setup}. 
We model the probe in water as a soft overdamped Brownian particle with diameter $d$, and the bacteria by soft active Brownian particles of the same diameter. 

The optical tweezers induce an approximately harmonic potential $U(\bm r) = k\bm r^2/2$ with stiffness $k$, and we assume that this potential is felt only by the passive probe. 
While optical tweezers are commonly applied to manipulate bacteria, the authors of Ref.~\cite{krishnamurthy2016micrometre} reduced the trapping of bacteria by the optical tweezers by adding glycerol to the solution, rendering our approximation reasonable.

We assume that the interactions between the probe and the bacteria, as well as between the individual bacteria, derive from the soft repulsive pair potential
\begin{equation}
    V(r) = 4\epsilon\left[ \left(\frac{d}{r}\right)^6 - \left(\frac{d}{r}\right)^3 + \frac{1}{4}\right]
\end{equation}
with energy scale $\epsilon$ and cutoff at the interparticle distance $r=2^{1/3}d$, i.e. $V(r>2^{1/3}d)=0$, providing the particles with the ``soft radius'' $d$. 

The probe position vector $\bm r$ thus obeys the Langevin equation
\begin{equation}
        \dot{\bm r} = \mu\bm F -\mu k\bm r + \sqrt{2D_t}\bm \xi\,,
\label{eq:eom-probe}
\end{equation}
and the dynamics of the ``bacterial'' position vectors ${\bm r}_i$ and orientations $\theta_i$, $i=1,\ldots,N$, are described by the Langevin equations
\begin{align}
    \begin{split}
        \dot{\bm r}_i &= \mu\bm F_i + v_a\bm n_i + \sqrt{2D_t}\bm \xi_i\,,\\
        \dot{\theta}_i &= \sqrt{2D_r}\nu_i\,.\\
    \end{split}
    \label{eq:eom-abp}
\end{align}
The repulsive forces read $\bm F = -\sum_i\nabla_{\bm r}V(|\bm r - \bm r_i|)$ and $\bm F_i = -\sum_j\nabla_{\bm r_i}V(|\bm r_i - \bm r_j|) - \nabla_{\bm r_i}V(|\bm r_i - \bm r|)$.
The probe and the active bath particles experience mutually independent unbiased Gaussian white noises $\bm \xi$ and ${\bm \xi_i}$ with translational diffusivity $D_t = \mu k_BT$, given by the temperature $T$, Boltzmann's constant $k_B$, and mobility $\mu $.
The interaction energy is set to $\epsilon=50k_BT_c$ with the minimum solvent temperature $T_c$.
For the active particles, we moreover need to account for the diffusion of their orientations $\bm n_i=(\cos\theta_i,\,\sin\theta_i)$, along which they self-propel at constant speed $v_a$, due to mutually independent unbiased Gaussian white noises $\nu_i$ with rotational diffusivity $D_r$. 

The dynamical equations are solved (iteratively) using Brownian dynamics simulations with time step $\mathrm{d}t = \SI{20}{\micro\second}$ inside a square box of side length $L$ with periodic boundary conditions.

\subsection{Parameters, units \& microscopic Stirling cycle}

We have chosen the parameters as close as possible to the experiment~\cite{krishnamurthy2016micrometre}.
We fixed the diameters of the probe and the bacteria to $d=\SI{1}{\micro\metre}$. 
(For a discussion of  the effect of choosing different sizes for the probe and bacteria, see Ref.~\cite{partII}.) 
The volume fraction $\phi = (N+1)\pi d^2/4L^2$ of the bacteria was kept constant at the relatively low value $\phi = 0.2$ to ensure that rare head-on collisions responsible for the large tails in the position distribution of the probe \cite{leptos2009dynamics-of-enhanced-tracer-diffusion, thiffeault2015distribution} were captured. 
All simulations were performed with $N = 100$ bacteria, corresponding to $L \approx \SI{19.8}{\micro\metre}$.
 
In the experiment~\cite{krishnamurthy2016micrometre}, the trap stiffness $k$, water temperature $T$, and bacterial activity, measured in our model by $v_a$ and $D_r$, were periodically modulated according to a protocol mimicking a Stirling cycle. 
Howere, the persistence time of the bacterial swimming motion reported in Ref.~\cite{krishnamurthy2016micrometre} was $\SI{1} {\second}$, regardless of other parameter values.
We therefore fixed the rotational diffusivity to $D_r=\SI{1}{\hertz}$, accordingly.
Consistently, since, in the experiment, the water temperature changed only moderately during the cycle, we set the particle mobility $\mu$ to a constant given by the Stokes value $\mu = \SI{5e7}{\second\per\kilo\gram}$ for a spherical Brownian particle with diameter $d=\SI{1}{\micro\metre}$ and the viscosity of water at $T_c=\SI{290}{\kelvin}$. 
For \emph{E. coli}, the swim speed was measured to be $\SI{35}{\micro\metre\per\second}$~\cite{berg1987rapid-rotation-of-flagellar-bundles}, which can be considered generic, since bacteria are generally reported to swim at around $\SI{50}{\micro\metre\per\second}$, or $10^2$ times their body length during one second~\cite{tortora2020microbiology}.
\textcolor{black}{Nevertheless, the swim speed is highly sensitive to the ambient temperature $T$, a characteristic that was experimentally exploited to switch the activity on and off (or nearly off).}
In terms of the non-dimensional P{\'e}clet number $\mathrm{Pe}=v_a/d D_r$, which gives a measure of the active transport during the orientational persistence time, this behavior is captured by the range of $\mathrm{Pe}\in[1,10^2]$ considered in our simulations.

\begin{figure}[t]
    \centering
    \includegraphics[width=\linewidth]{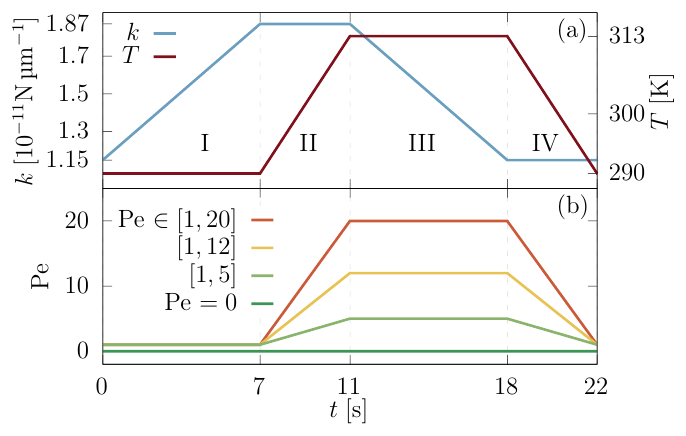}
    \caption{Periodic protocols for the trap stiffness $k(t)$ and solvent temperature $T(t)$ in (a) and the bath particles' P{\' e}clet number $\mathrm{Pe}(t)$ in (b), as employed in the simulations.
    }
    \label{fig:protocol}
\end{figure}

In Fig.~\ref{fig:protocol}(a), we depict the protocols for stiffness and temperature employed in the experiments, as well as in our simulations. In Fig.~\ref{fig:protocol}(b), we additionally show the protocol used for $v_a$ in the simulation, which, we believe, captures the main features of the protocol used in Ref.~\cite{krishnamurthy2016micrometre}. 
The experimentally implemented Stirling cycle consists of four linear branches. 
During ``isothermal compression'' (I), the trap stiffness $k$ increases from $k_{\min} = \SI{1.15e-11}{\newton\per\micro\metre}$ to $k_{\max} = \SI{1.87e-11}{\newton\per\micro\metre}$ while the water temperature $T=T_c= \SI{290}{\kelvin}$ and $\mathrm{Pe}$ take their minimum values.
During the subsequent isochoric heating (II), the stiffness is kept constant at $k=k_{\max}$, and $T$ increases from $T_c$ to $T_h= \SI{313}{\kelvin}$ and the bacterial swimming intensifies from its minimum to its maximum. 
The following isothermal compression (III) is achieved by the decrease of the stiffness from  $k_{\max}$ to $k_{\min}$ at $T=T_h$ and maximum $\mathrm{Pe}$.
The cycle is closed by an isochoric cooling, where the temperature and P{\'e}clet number return to their initial (minimum) values at constant stiffness $k_{\min}$ (IV). 
Its total duration is $t_{\mathrm p}=\SI{22}{\second}$ and those of the isothermal and isochoric processes are $\SI{7}{\second}$ and $\SI{4}{\second}$, respectively, all in accord with the experiment.
Importantly, the relaxation time $1/\mu k$ of the probe in the trap ranges from $1/\mu k_{\max}\approx\SI{1.1}{\milli\second}$ to $1/\mu k_{\min}\approx\SI{1.74}{\milli\second}$, so that the engine operation can be considered quasi-static. 
It also always remains below the sampling rate $\Delta t=\SI{2}{\milli\second}$ used in our simulations and experiments~\cite{krishnamurthy2016micrometre} to save numerical observables.

As indicated in the plots, length is measured in units of $d=\SI{1}{\micro\metre}$, time in units of $1/D_r=\SI{1}{\second}$ and energy in units of $k_BT_c=\SI{4.002}{\newton\micro\metre}$.

\section{Theoretical analysis}
\label{sec:theory}

In this section, we first review stochastic thermodynamic definitions of work, heat, and efficiency. 
Then we discuss how the notion of thermodynamic temperature can be extended to heat engines operating with non-equilibrium reservoirs, like in the experiment under consideration. 
Finally, we present analytical results for two equivalent versions of an effective model that maps the whole setup onto a single swimmer in a trap: 
the active Brownian heat engine (ABE) based on an individual active Brownian or active Ohrstein-Uhlenbeck particle~\cite{holubec2020active-brownian-heat-engines}, respectively.
Both versions serve as coarse-grained descriptions of the many-body simulations and yield consistent energetics, although they differ with respect to their stochastic fluctuations, as further elaborated in Ref.~\cite{partII}.

\subsection{Work, heat and efficiency}
\label{sec:IIIA}

The efficiency of a (heat) engine is defined as the ratio of the produced (average) output work and the total supplied energy input (heat),
\begin{equation}
    \eta = \frac{W_{\text{out}}}{Q_{\text{in}}}\,. 
    \label{eq:eta}
\end{equation}
In stochastic thermodynamics, a system's known energy function and its dependency on the control parameters of the driving protocol can be utilized to define work and heat~\cite{sekimoto2010stochastic-energetics, seifert2012stochastic-thermodynamics}.
In our case, the total internal energy of the probe with mass $m$ is 
\begin{equation}
    U = \frac{k \sigma}{2} + \frac{m\langle{\bm v}^2\rangle}{2}\,.
\label{eq:energy}
\end{equation}
The first term $k \sigma/2 = k\langle{\bm r}^2\rangle/2$ is the ensemble-averaged potential energy of the probe in the harmonic potential, the second term is the probe's average kinetic energy. 
Since the particle's momentum 
$m {\bm v}$ plays no role in the overdamped dynamics of the probe, the kinetic energy is often neglected when assessing the efficiency of stochastic heat engines. Although this omission can lead to overestimations, as discussed in~\cite{Dechant_2017}, we show later that it is of minor relevance to the present discussion.

In accordance with standard procedures of stochastic thermodynamics~\cite{sekimoto2010stochastic-energetics,seifert2012stochastic-thermodynamics} and also with the experiment~\cite{krishnamurthy2016micrometre}, we define work via the contribution $\dot{k} \sigma/2$ to the rate of change $\dot{U}$ of the internal potential energy $U$ due to the external driving of the system in $k$. 
The remainder $k \dot{\sigma}/2 + m \langle\bm v\cdot\dot{\bm{v}}\rangle$ of $\dot{U}$ is then the energy input per time from the reservoir to the confined probe (the ``working medium''), which we identify as the heat current. 
\textcolor{black}{We note as an aside that our energy balance does not account for the power $\dot{Q}_{\rm hk}$, which is technically necessary to keep the bath active (often called housekeeping heat)~\cite{holubec2020active-brownian-heat-engines, malgaretti2021work-cycle, ekeh2020thermodynamic-cycles, pietzonka2022second-law-for-active-heat-engines,Oh2023}, but only for the heat acquired from this active bath by the working substance}. 

The total output work $W_{\text{out}}=-W$ and input heat $Q_{\text{in}}$ per cycle follow from integrating the energy balance over time $t=0\ldots t_p$.
For the net output work, we get without any ambiguity the expression 
\begin{equation}
    W_{\text{out}} = - \frac{1}{2}\int_0^{t_p} \dot{k}(t)\sigma(t)\,\mathrm dt\,.
\label{eq:wout}
\end{equation}

For the input heat, the situation is more complicated since it admits several definitions, as will be discussed in the following.
The positive energy transferred from the bath is
\begin{equation}
    Q_{\text{in}} = \frac{1}{2}\int_0^{t_p} k(t)\dot{\sigma}(t)\Theta(\dot{\sigma})\,\mathrm dt\,,
 \label{eq:heatOverFull}
\end{equation}
where the integrand is nonzero during branches II and III of the cycle in Fig.~\ref{fig:protocol} and only heating but no cooling of the probe is counted, as ensured by the Heaviside step function $\Theta(\dot{\sigma})$. 
\textcolor{black}{
This is the most commonly employed definition of input heat in studying overdamped stochastic heat engines, both in theory~\cite{seifert2012stochastic-thermodynamics,holubec2020active-brownian-heat-engines} and experiment~\cite{blickle2012realization}.
}

To minimize avoidable losses and enhance their efficieny, macroscopic Stirling engines can be operated using a recuperation mechanism, which stores the heat leaving the working medium along the cooling branch IV and returns it during the heating branch II. 
For perfect recuperation, only the input heat $Q_{\text{in}}^\star$ transferred from the bath during the isothermal branch III with $T(t)\equiv T_h$ counts as a net cost, or, using the Kronecker symbol $\delta$, the branch with $\delta_{T_h, T(t)} = 1$:
\begin{equation}
    Q_{\text{in}}^\star = \frac{1}{2}\int_0^{t_p}k(t)\dot{\sigma}(t) \delta_{T_h, T(t)}\,\mathrm dt\,.
    \label{eq:Qstar}
\end{equation}
While this definition is quite useful for classical designs, it is hard to imagine what a recuperation mechanism storing the energy released by the ``working medium'' to the active bath should look like.

In any case, the above expressions do not account for the rate of kinetic energy change $m \dot{\langle{\bm v}^2\rangle}/2$.
In the overdamped approximation, momentum degrees of freedom of the particle are assumed to be in thermal equilibrium with a background solvent at all times.
Thus $m \dot{\langle{\bm v}^2\rangle}/2 =  k_B \dot{T}$, and the kinetic contribution to $Q_{\text{in}}$ is $k_B(T_h-T_c)$, while it vanishes for $Q_{\text{in}}^\star$.

For conventional quasi-static Stirling cycles such as the one depicted in Fig.~\ref{fig:protocol}(a) with $\mathrm{Pe}(t)=0$, the most commonly used expression~\eqref{eq:heatOverFull} for the heat supply yields~\cite{blickle2012realization}
\begin{equation}
    \eta_{\infty}^{\text{eq}} = \eta_{\text{C}} \left[1 + \frac{\eta_{\text{C}}}{\log\frac{k_{\max}}{k_{\min}}}\right]^{-1}
    \label{eq:eta-inf-eq}
\end{equation}
for the efficiency from Eq.~\eqref{eq:eta}.
Taking into account the kinetic contribution to the input heat, the latter changes to $W_{\text{out}}/(Q_{\text{in}} + k_B(T_h-T_c))$, which is always smaller, but the correction becomes negligible for active engines operating with a large activity difference, as shown in Fig.~\ref{fig:eta-vs-Teff}(b), which is pertinent to the experimental setup.

Assuming perfect recuperation, the input heat is given by Eq.~\eqref{eq:Qstar}, no matter whether the kinetic contribution to heat is counted or not, and the efficiency reads
\begin{equation}
    \eta^\star = \frac{W_{\text{out}}}{Q_{\text{in}}^\star}\,.
    \label{eq:eta-star}
\end{equation}
For quasi-static passive engines, $\eta^\star$ is then equivalent to the Carnot efficiency, $\eta_{\text{C}}=1-T_c/T_h$, the optimum achievable for any heat engine, independently of design and working substance, and independently of how the heat bath is practically realized. 

\textcolor{black}{In their methods section, the authors of Ref.~\cite{krishnamurthy2016micrometre} define the efficiency of the bacterial heat engine as $\eta^*$. However, the discussion in the main body of their paper, where the obtained efficiencies are compared to the upper bound on Stirling efficiency \eqref{eq:eta-inf-eq} valid for $\eta = W_{\text{out}}/Q_{\text{in}}$, seems to align more closely with the standard efficiency definition \eqref{eq:eta}. 
This suggests that the authors might have intended to use the efficiency  $\eta$ instead of $\eta^*$.}

Besides the standard efficiencies \eqref{eq:eta} and \eqref{eq:eta-star}, defined as average work over average heat, one may also consider the stochastic efficiency $\eta_s = -w/q_{\text{in}}$, where $w$ and $q_{\text{in}}$ are stochastic work and heat per cycle, defined in App.~\ref{app:eta}. 
The large deviation function of $\eta_s$ does provide useful information about the engine efficiency \cite{proesmans2015stochastic-efficiency, verley2014unlikely-carnot-efficiency, martinez2015brownian-carnot-engine,verley2014efficiency-fluctuations, gingrich2014efficiency-and-large-deviations}, but only when very large sample sizes are available.
However, $\eta_s$ itself is not so useful for the study of average thermodynamics because all its moments diverge~\cite{esposito2015efficiency-statistics-at-all-times}.
Yet, it is sometimes still being employed for that purpose \cite{Kwon2024}.
In Fig.~\ref{fig:SM-eta-average} in App.~\ref{app:eta}, we plot the running mean of $\eta_s$ over the number of cycles to illustrate the fact that $\eta_s$ is not a well-behaving stochastic quantity upon averaging. 

\textcolor{black}{
As mentioned above, none of the above definitions of input heat considers the power $\dot{Q}_{\rm hk}$ that must be constantly supplied into the bath to sustain its activity or other details of the active bath dynamics. 
The above-defined thermodynamic efficiencies thus measure how effectively the engine transforms the energy acquired from the bath, irrespective of how the bath is kept active.
They solely depend on the position variance of the probe and the trap stiffness protocol, which can, for the given class of colloidal engines with an active bath, be understood as the relevant thermodynamic degrees of freedom. 
This is why we refer to the above-defined efficiencies as thermodynamic efficiencies.
Specific technical details affecting the variance—whether through clustering, persistence, increased effective diffusion, or details of probe-bath interaction—are therefore implicitly included in the thermodynamic analysis and do not need to be resolved, at this stage.
A notable difference between the present situation and classical thermodynamics with equilibrium heat baths is that alterations of the probe-particle interactions or other changes in the bath dynamics can affect the engine performance via their effect on the position variance in ways that would not be possible for equilibrium baths.}

\textcolor{black}{
In specific situations when the housekeeping power $\dot{Q}_{\rm hk}$ is known for a given setup, the thermodynamic efficiencies $\eta$ and $\eta^*$ can be used to calculate the `technical' efficiency of that specific technical realization of the engine, $\eta_{\rm t}$. The nominal
energy input per cycle is increased by a housekeeping contribution $t_p \dot{Q}_{\rm hk}$ that accounts for the corresponding `technical losses'. Then, one writes for the technical efficiency
\begin{equation}
\eta_{\rm t}=  \frac{W_{\rm out}}{Q_{\rm in} + t_p \dot{Q}_{\rm hk}}= \frac{\eta Q_{\rm in}}{Q_{\rm in} + t_p \dot{Q}_{\rm hk}}
\label{eq:eta_tot_full}
\end{equation}
and similarly for $\eta^*$. The housekeeping heat $ t_p \dot{Q}_{\rm hk}$ either has to be measured or theoretically estimated by specific thermodynamically consistent models for active baths~\cite{Pietzonka2017,ryabov2022enhanced,Fritz2023}. Similarly, one can also consider the power required to realize the harmonic potential using optical tweezers and other energy sources, needed for a practical realization of the idealized setup in the lab. In most realistic situations, such technical efficiencies accounting for auxiliary operational costs will be very small.}

\subsection{Effective temperature}
Thermodynamic temperature is classically defined by the zeroth and second laws of thermodynamics \cite{callen1985thermodynamics}.
Since the zeroth law relies on mutual equilibration, it does not hold for out-of-equilibrium systems such as active systems.
For these, each degree of freedom (or thermometer) $i$ may acquire a different (effective) temperature $T_{\text{eff}}^i$ \cite{zakine2017stochastic-stirling, baldovin2017about-thermometers, puglisi2021optimization}, which merely needs to be consistent with the second law, in sharp contrast to equilibrium where all thermometers (no matter which degree of freedom they couple to) must show identical results. 
Breaking the zeroth law is thus what allows active heat engines to operate at meaningful effective bath temperatures $T_{\text{eff}}\sim\mathcal O(\SI{e3}{\kelvin})$ and thereby achieve large efficiencies, approaching the limit $\eta=1$ of work-to-work conversion, without burning the laboratory.

Even though there is an ongoing debate in the literature, which effective temperature is the most suitable for assessing the thermodynamic performance of an active engine~\cite{zakine2017stochastic-stirling,holubec2020active-brownian-heat-engines, park2020active-reservoirs,puglisi2021optimization, chang2023stochastic-heat-engines}, we demonstrate below that it should be the effective temperature linked to the degrees of freedom, which act as the engine's working medium. 
Only a correspondingly defined effective temperature provides, if it exists, a proper second law upper bound on the engine efficiency and thereby a consistent notion of heat and heat engine (as opposed to imperfect work-to-work transducers, say)~\cite{holubec2020active-brownian-heat-engines}.  
An appropriate notion of effective temperature is not guaranteed to exist when the working medium comprises more than a single degree of freedom, as different degrees of freedom can have different $T_{\text{eff}}$~\cite{holubec2020active-brownian-heat-engines}. 
For example, when using the work and heat definitions 
\eqref{eq:wout} and \eqref{eq:heatOverFull} (neglecting kinetic contributions to heat, for simplicity) or \eqref{eq:Qstar}, we find that the work-producing degree of freedom of the medium is the position variance $\sigma$, and thus $T_{\text{eff}}$ exists. 
On the other hand, using a more elaborate (``underdamped'') heat definition by extending Eq.~\eqref{eq:heatOverFull} to account for the kinetic term from Eq.~\eqref{eq:energy} means that the working medium is understood to comprise both position and momentum variances, and the existence of $T_{\text{eff}}$ is not generally guaranteed~\cite{holubec2020underdamped}.

Clearly, if the effective temperature does not exist, it does not make sense, thermodynamically, to speak of a heat engine.
The reason is that it then becomes impossible to determine, with thermodynamic means, how much of the energy interchanged between the working fluid and the bath is ``disordered heat'' or ``work in disguise''.
For thermodynamic cycles with equilibrium heat reservoirs, temperatures $T_i$ relate the energies $Q_i$ extracted from the individual reservoirs to their entropy change $\Delta S_i = - Q_i/T_i$, via the Clausius equality.
Since the entropy change of the working medium per cycle vanishes, the temperatures $T_i$ provide a mapping between the energy exchanges $Q_i$ with the baths and the total entropy change
per cycle, $\Delta S_{\rm tot} = -\sum_i Q_i/T_i$.
The second law $\Delta S_{\rm tot} \ge 0$ then implies an upper bound $\eta < 1$ on the engine's efficiency, which saturates for quasi-static reversible cycles. 
For cycles with non-equilibrium reservoirs, no such \emph{a priori} upper bound on efficiency exists unless one can assign a set of effective temperatures $T_i^{\rm eff}$ to the non-equilibrium baths, providing a thermodynamically consistent replacement for the above $T_i$.
If such $T_i^{\rm eff}$ do not exist, i.e., one cannot relate the energy transferred from the reservoirs with a thermodynamically consistent entropy production, no second law bound on efficiency exists, and one can only apply the first law limitation $\eta \le 1$, also valid for work-to-work conversion.

For quasi-static processes, when a system is in contact with non-equilibrium reservoirs, effective temperatures of individual degrees of freedom can be assigned by comparing the observed behavior with quasi-equilibrium behavior under the same protocol while in contact with equilibrium reservoirs at well-defined temperatures. 
Consider, for example, a system described by a Hamiltonian $H(k,\bm r)=k f(\bm r) / 2$, where $k(t)$ is the externally modulated parameter \textcolor{black}{and $f(\bm r)$ is an arbitrary function of particle's position with finite mean}. 
In this case, the working medium has the single degree of freedom $\sigma = \langle f(\bm r) \rangle$, work and heat currents are given by $\dot{k} \sigma / 2$ and $k \dot{\sigma} / 2$, and $T_{\rm eff}$ exists.
Under equilibrium conditions, $\langle f(\bm r) \rangle_{\text{eq}} = \sigma_{\text{eq}}(k,T) = \int f(\bm r)e^{-H(k,\bm r) /k_BT}/Z(k,T)$, where $Z(k,T)$ is the partition function.
The effective temperature $T_{\text{eff}}$ can thus be calculated from the implicit equation $\sigma_{\text{eq}}(k,T_{\rm eff}) = \sigma$. 
It corresponds to the temperature of an equilibrium bath that, if used as a replacement for the non-equilibrium bath, would give the same thermodynamic performance.
Since the assigned $T_{\rm eff}$ thus follows from the Boltzmann distribution, its thermodynamic consistency with the second law is guaranteed by construction. 
Noteworthily, identical working media in contact with different non-equilibrium baths at the same $T_{\text{eff}}$ yield the same thermodynamic performance, which provides a notion of (reduced) thermodynamic universality in the absence of the zeroth law.

The work and heat definitions in Eqns.~\eqref{eq:wout} and \eqref{eq:heatOverFull} (neglecting kinetic contribution to heat) or \eqref{eq:Qstar} are related to the above general discussion via the function $f(\bm r)={\bm r}^2$.
In two dimensions, $k\sigma_{\text{eq}}/2= k_BT$ and thus 
\begin{equation}
    T_{\text{eff}} = \frac{k\sigma}{2k_B} \,.
    \label{eq:Teff-var}
\end{equation}

\subsection{Results for the active Brownian heat engine}

In the numerical model in Sec.~\ref{sec:model}, the effect of the active bath on the probe is described by the combined active (bacterial) and passive (solvent) noise ${\bm\eta} \equiv \mu\bm F + \sqrt{2D_t}\bm \xi$ in Eq.~\eqref{eq:eom-abp}, with which the dynamical equation for the probe can be rewritten in the form 
\begin{equation}
\dot{\bm r} = -\mu k\bm r +\bm\eta.
\label{eq:ABE}
\end{equation}
According to the above discussion, identical working media in contact with different non-equilibrium baths yield the same thermodynamic performance if the baths are described by the same $T_{\text{eff}}$. 
This element of universality far from equilibrium implies, in particular, that the complex non-equilibrium many-body problem posed by the operation of the active heat engine can be mapped onto a much simpler toy model involving a single active particle, namely an animated probe, self-propelled by some effective noise $\bm\eta$.
By finding the effective temperature corresponding to Eq.~\eqref{eq:ABE} for a given active-particle model, one can thus develop a toolkit of analytically tractable toy models for the original bacterial heat engine and compute analytical predictions for its thermodynamic performance.

Arguably, the simplest active noise $\bm\eta$ in Eq.~\eqref{eq:ABE}, which genuinely brings the trapped colloid out of equilibrium, is exponentially correlated noise, such as in the active Brownian particle (ABP)~\cite{holubec2020active-brownian-heat-engines} or active Ornstein-Uhlenbeck particle (AOUP)~\cite{caprini2022parental-active-model} models \textcolor{black}{(for more details concerning these two models, see App.~\ref{apx:ABPAOUP})}. 
More precisely, the $\bm\eta$-autocorrelation function corresponding to the ABP and AOUP models can be written in the form $\langle\eta_i(t) \eta_j(t')\rangle = \delta_{ij}v_a^2 e^{-D_r|t-t'|} + 2D_t\delta_{ij}\delta(t-t')$, and thus $\bm\eta$ comprises an exponentially correlated component, representing persistent motion due to activity, and an independent white component, describing thermal background noise. 

For both models the quasi-static variance $\sigma_{\infty}$ reads~\cite{holubec2020active-brownian-heat-engines}
\begin{equation}
    \sigma_{\infty}(k,T,v_a) = \frac{2 k_B T_{\text{eff}}}{k} = \frac{2k_BT}{k} + \frac{v_a^2}{\mu k(\mu k + D_r)}\,,
    \label{eq:sigma2-inf}
\end{equation}
so that they predict the same energetics and thus the same thermodynamic performance. 
The two models differ in the probability density of the noise, though. 
While it is non-Gaussian for the ABP, it is Gaussian for the AOUP. 
In a certain parameter regime, the probe's noise $\bm\eta$, measured in our many-body simulations, becomes Gaussian and exponentially correlated, and, in this regime, the AOUP also reproduces fluctuations of work and heat very satisfactorily~\cite{partII}, but this is inconsequential for the present discussion.

\textcolor{black}{As explained in the previous Sec.~\ref{sec:IIIA}, all heat engines with the same position variance under the given stiffness protocol are thermodynamically equivalent in the sense that their efficiency and output work, as defined above, are identical. We now follow Ref.~\cite{holubec2020active-brownian-heat-engines} in summarizing the class of thermodynamically equivalent engines with exponentially correlated noise in Eq.~\eqref{eq:ABE} and the quasistatic effective temperature $T_{\text{eff}}$ given by Eq.~\eqref{eq:sigma2-inf}, using the notion of active Brownian heat engine (ABE).
}
For the ABE class, the quasi-static thermodynamic performance can be investigated analytically by substituting its prediction for the quasi-static variance $\sigma_{\infty}$ in Eq.~\eqref{eq:sigma2-inf} into the definitions of work and heat in Eqs.~\eqref{eq:wout}--\eqref{eq:Qstar}.
Work is performed only during the expansion and compression branches, i.e., $W_{\text{out}}= -W_{\text{I}} -W_{\text{III}}$,
\begin{align}
    W_{\text{I}} &= \int_{k_{\min}}^{k_{\max}}\sigma_{\infty}(T_c,v_{\min})\,\mathrm dk\,,\\
    W_{\text{III}} &= \int_{k_{\max}}^{k_{\min}}\sigma_{\infty}(T_h, v_{\max})\,\mathrm dk\,.
\end{align}
The heat supplied by the hot active bath during branch III is given by the first law, $Q_{\text{in}}^{\star}=\Delta U_{\text{III}} - W_{\text{III}}$, with the energy difference 
\begin{equation}
    \Delta U_{\text{III}} = U(k_{\min}, T_h, v_{\max}) - U(k_{\max}, T_h, v_{\max}),
\end{equation}
where $U(k,T,v_a) = k \sigma_\infty(k,T,v_a)/2$.

The total heat input~\eqref{eq:heatOverFull} is $Q_{\text{in}}=\Delta U_{\text{II}} + \Delta U_{\text{III}} - W_{\text{III}}$, where
\begin{equation}
    \Delta U_{\text{II}} = U(k_{\max}, T_h, v_{\max}) - U(k_{\max}, T_c, 0)
\end{equation}
and the complete expressions for the efficiencies $\eta_{\infty}^\star=W_{\text{out}}/Q_{\text{in}}^\star$ and  $\eta_{\infty} = W_{\text{out}}/Q_{\text{in}}$ are then,
\begin{widetext}
    \begin{equation}
        \label{eq:eta-inf-star}
        \eta_{\infty}^\star =\frac{(T_h-T_c)\log\frac{k_{\max}}{k_{\min}} + \frac{\mu k_{\min} + D_r}{D_r} \left[\Delta T_{\text{eff}} - (T_h-T_c)\right] \log\frac{k_{\max}(\mu k_{\min} + D_r)}{k_{\min}(\mu k_{\max} + Dr)} }{ T_h \log\frac{k_{\max}}{k_{\min}} + (\mu k_{\min}+D_r) \left[\Delta T_{\text{eff}} - (T_h-T_c)\right]
                 \left[ \frac{\mu(k_{\max} - k_{\min})}{(\mu k_{\min} + D_r)(\mu k_{\max} + D_r)} + \frac{1}{D_r} \log \frac{k_{\max}(\mu k_{\min} + D_r)}{k_{\min}(\mu k_{\max} + D_r)} \right]}\,,
    \end{equation}
    \begin{equation}
        \label{eq:eta-inf}
        \eta_{\infty} = \frac{(T_h-T_c)\log\frac{k_{\max}}{k_{\min}} + \frac{\mu k_{\min} + D_r}{D_r}\left[ \Delta T_{\text{eff}} - (T_h-T_c)\right] \log\frac{k_{\max}(\mu k_{\min} + D_r)}{k_{\min}(\mu k_{\max} + D_r)}}
                    {(T_h-T_c) + T_h\log\frac{k_{\max}}{k_{\min}} + \left[ \Delta T_{\text{eff}} - (T_h-T_c) \right] \left[ 1 + \frac{\mu k_{\min} + D_r}{D_r}\log\frac{k_{\max}(\mu k_{\min}+D_r)}{k_{\min}(\mu k_{\max} + D_r)}\right]}\,.
    \end{equation}
\end{widetext}
Here, the effective temperature difference $\Delta T_{\text{eff}} = \max  \Delta T_{\text{eff}} - \min  \Delta T_{\text{eff}}$ is
\begin{equation}
    \Delta T_{\text{eff}} = T_h - T_c + \frac{v_{\max}^2}{2\mu k_B(\mu k_{\min} + D_r)}.
    \label{eq:DTeff}
\end{equation}
Note that the minimum swim speed is assumed to be zero, $v_{\min}=0$, to simplify the expression and reduce the number of parameters.
Small minimum swim speeds of $v_a\approx\SI{1}{\micro\metre\per\second}$ amount to small changes in the effective temperature $\Delta T_{\rm eff}\approx\SI{1}{\kelvin}$ via Eq.~\eqref{eq:sigma2-inf} and become negligible for the large temperature differences considered in the following section (and probably in most meaningful applications).

In the limit of a passive cycle, i.e., $v_{\max}\to 0$ and $\Delta T_{\text{eff}}\to T_h - T_c$, $\eta_{\infty}^\star$ reduces to $\eta_{\text{C}}$ and $\eta_{\infty}$ to the conventional quasi-static Stirling efficiency~\eqref{eq:eta-inf-eq}.
The same limits are reached for vanishing persistence, $D_r\to\infty$, when both the ABP and AOUP reduce to a passive Brownian particle.
In the opposite limit of infinite persistence, i.e., $v_{\max}\to\infty$ and $\Delta T_{\text{eff}}\to\infty$, one obtains well-defined limits that are distinct from their conventional counterparts,
\begin{align}
    \lim_{\Delta T_{\text{eff}}\to\infty}\eta_{\infty} &= \left[ 1 + \frac{\frac{D_r}{\mu k_{\min} + D_r}}{\log\frac{k_{\max}(\mu k_{\min} + D_r)}{k_{\min}(\mu k_{\max} + D_r)}}\right]^{-1}\,,\\
    \lim_{\Delta T_{\text{eff}}\to\infty}\eta_{\infty}^\star &=
    \frac{ \log \frac{k_{\max}(\mu k_{\min} + D_r)}{k_{\min}(\mu k_{\max} + D_r)}}{  \log \frac{k_{\max}(\mu k_{\min} + D_r)}{k_{\min}(\mu k_{\max} + D_r)} + \frac{D_r\mu(k_{\max} - k_{\min})}{(\mu k_{\min} + D_r)(\mu k_{\max} + D_r)}}\,.
\end{align}
Notably, for all $\Delta T_{\text{eff}}> T_h-T_c$, the relations $\eta_{\text{C}}>\eta_{\infty}^\star$ and $\eta_{\infty}^{\text{eq}}>\eta_{\infty}$ hold. 
This is caused by the stiffness dependence of the effective temperature in Eq.~\eqref{eq:sigma2-inf}.
It originates from the interference of the potential with the persistent motion of the active bath particles and implies that the active engine implemented in Ref.~\cite{krishnamurthy2016micrometre} does not follow a Stirling protocol in the $T_{\text{eff}}-k$ diagram, even under quasi-static driving, as depicted in Fig.~\ref{fig:var-t-and-Teff-k}. 
It therefore has a lower efficiency than a Carnot engine with a ``passive'' equilibrium heat bath, which does not induce such deviations.
For finite $v_{\max}$, the efficiency is determined by the competition of the time scales $1/\mu k$ and $1/D_r$. 
The studied model operates in the regime where the persistence time dominates over the equilibration time in the harmonic trap, $1/D_r\gg 1/\mu k$, but other choices could be worthwhile to explore experimentally. 

As shown next, the above results offer a fit-free analytical description, in terms of $\Delta T_{\rm eff}$, of the efficiencies obtained from our many-body simulations for the parameter values provided in Sec.~\ref{sec:model}.

\section{Simulation results}
\label{sec:results}

\begin{figure}[t]
    \centering
    \includegraphics[width=\linewidth]{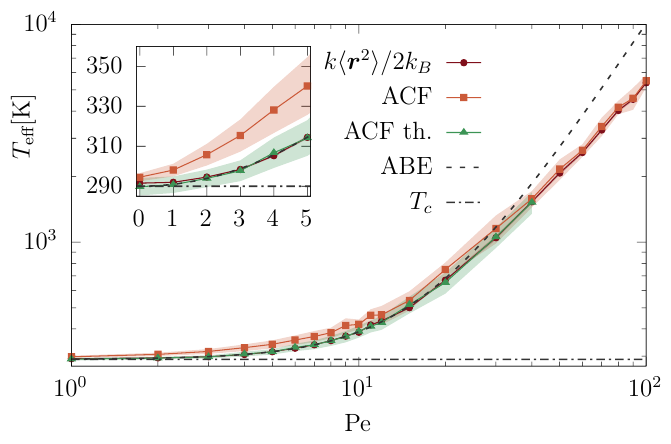}
    \caption{The effective temperature $T_{\text{eff}}$ as function of the P{\'e}clet number, quantifying the activity.
    Many-body simulations of a passive probe in an active bath (symbols) are well parametrized by the effective ABE (dashed line), based on a single active particle in the same trap and background solvent.
    Different symbols correspond to the evaluation of $T_{\rm eff}$ from the probe's position variance (dark red circles) and the autocorrelation functions of its driving noise (green triangles). \textcolor{black}{Red squares represent the effective temperature obtained from the autocorrelation function, considering only the athermal (active) component of the noise, $\mu \bm{F}$, in Eq.~\eqref{eq:Teff-acf}. This effective temperature corresponds to the position variance in the limit where the intensity of the background equilibrium fluid vanishes ($D_t = 0$) \cite{holubec2020active-brownian-heat-engines}.} 
    All effective temperatures were obtained at constant solvent temperature $T=T_c$ (dash-dotted baseline) and trap stiffness $k=k_{\min}$.
    }
    \label{fig:Teff-methods}
\end{figure}

In this section, we describe the main numerical results obtained by solving our implementation of the bacterial engine of Ref.~\cite{krishnamurthy2016micrometre}, as described in Sec.~\ref{sec:model}, using Brownian dynamics simulations.

In Fig.~\ref{fig:Teff-methods}, the symbols represent the effective temperature $T_{\text{eff}} = k \sigma/(2 k_B)$ according to Eq.~\eqref{eq:Teff-var} as a function of the P{\'e}clet number $\mathrm{Pe}~=~v_a/d D_r$, for fixed solvent temperature $T$ and stiffness $k$.
We determined the position variance from the simulation data using two complementary approaches.
First, we directly averaged the squared magnitude of the position vector (dark red circles).
Secondly, we calculated $\sigma$ from the autocorrelation function of the active-bath noise $\bm \eta = \mu\bm F + \sqrt{2D_t}\bm \xi$ as detailed in App.~\ref{app:acf} (green triangles).
\textcolor{black}{
Both procedures yield consistent results.
Red squares illustrate that neglecting the thermal solvent noise $\bm\xi$ contribution to $\bm \eta$ only leads to a slight overestimation of $T_{\rm eff}$, showing that the active and thermal noise contributions are relatively strongly anti-correlated for low activities ($\mathrm{Pe}< 50$).
(The thermal noise disturbs the active swimming slightly, decreasing the effective temperature of the active bath compared to the situation with active particles in a deterministic solvent.) 
For stronger activities, the ACFs with and without the thermal noise overlap.}
The dashed line in Fig.~\ref{fig:Teff-methods} shows $T_{\text{eff}}$ from Eq.~\eqref{eq:sigma2-inf} for the much simpler schematic one-particle ABE model, evaluated using the parameters from Sec.~\ref{sec:model},    except for the swim speed $v_a = \SI{9}{\micro m/s}$, which was adjusted to fit the simulation results at $\mathrm{Pe}=10$ ($v_a = \SI{10}{\micro m/s}$).
\textcolor{black}{
The comparison between the fit and our numerical results from the many-body simulations shows that the passive probe’s motion closely resembles that of the active bath particles, albeit with a slightly lower overall speed. This suggests that the transfer of activity from the bath particles to the probe is good but not perfect. Indeed, non-central collisions and, at higher densities, competing collision forces from multiple directions are all expected to limit the activity transfer in our “dry” swimmer model, which does not account for more realistic, yet intricate and varied hydrodynamic (and other, non-reciprocal~\cite{Auschra2021}) interactions between the microswimmers and the colloidal probe. The observed speed mismatch becomes somewhat more pronounced at higher Péclet numbers, consistent with a moderate clustering of bath particles around the probe~\cite{partII}, effectively enlarging its radius and friction.} 
Although the numerical data for $T_{\rm eff}$ are therefore not perfectly parabolic in $\mathrm{Pe}$ at higher values, the ABE prediction, Eq.~\eqref{eq:sigma2-inf}, nicely agrees with the simulation up to $T_{\text{eff}}\approx\SI{e3}{\kelvin}$. Punctual agreement between the two models at higher $T_{\rm eff}$ can, of course, be enforced by fitting the ABE's $v_a$ to the measured $\sigma$, at the desired $\mathrm{Pe}$ value. 
In Ref.~\cite{partII}, we further show that $T_{\rm eff}$ increases quadratically with the probe radius, in agreement with Eq.~\eqref{eq:sigma2-inf} for $D_r\ll \mu k$ (as experimentally the case) since the mobility $\mu$ is inversely proportional to the radius.

\begin{figure}
    \centering
    \includegraphics[width=\linewidth]{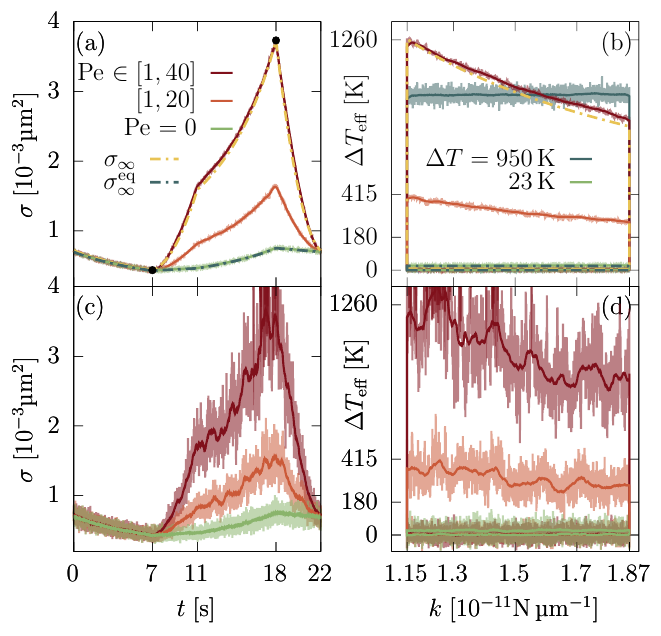}
    \caption{
    Simulation results and theoretical predictions from the ABE model for the Stirling protocol of Fig.~\ref{fig:protocol}.
    Left panels: position variance $\sigma$ as a function of time.
    Right panels: phase portraits depicting the effective temperature difference $\Delta T_{\rm eff} = T_{\rm eff} - \min T_{\rm eff}$ against the trap stiffness $k$.
    Transparent lines were obtained by calculating $\sigma$ as an average over several thousand (upper panels) and hundred (lower panels) consecutive cycles. 
    The solid lines are moving averages of the transparent lines over 200 time points. 
    In (a) and (b), the simulation data is complemented by analytical results obtained from the ABE model, Eq.~\eqref{eq:sigma2-inf}, for the passive case and for $\mathrm{Pe} \in [1,40]$ (blue and yellow dashed lines, respectively). 
    The minimum and maximum values of $v_a$ in the active ABE were determined by adjusting the variance of the ABE to the simulations at the two black dots in (a).
    In (b) a conventional Stirling cycle (blue line), operating with a passive bath at the extreme temperature difference $\Delta T=\SI{950}{\kelvin}$, is shown to underscore the difference between the protocol that the original experiments attempted to implement and the actual cycle of the active engine.
    }
    \label{fig:var-t-and-Teff-k}
\end{figure}

Our first main result is Fig.~\ref{fig:var-t-and-Teff-k}, showing the time evolution of the position variances (left panels) and the dependence of the corresponding effective temperatures on the trap stiffness (right panels) during the cycle depicted in Fig.~\ref{fig:protocol}, for several activity protocols. 
The data in the upper panels were generated from several thousands of consecutive cycles to obtain relatively smooth lines.
In contrast, the lower panels mimic the data acquisition in the experiment, with only a hundred consecutive cycles~\cite{krishnamurthy2016micrometre}. 
The figure demonstrates that despite the relatively poor statistics attainable in experiments, a quantitative analysis of the engine's performance using $T_{\text{eff}}$ is still possible.
As discussed in the previous section, Figs.~\ref{fig:protocol} and~\ref{fig:var-t-and-Teff-k} contain all the information necessary to assess the average thermodynamic performance of the engine.

The shapes of the cycles in Fig.~\ref{fig:var-t-and-Teff-k}(b) and (d) show that only for a passive bath ($\mathrm{Pe}=0$) the engine operates along a proper Stirling cycle, which corresponds to a rectangle in the $T_{\text{eff}}-k$ plane.
Otherwise ($\mathrm{Pe}>0$), the effective temperature during the expansion branch of the cycle (III in Fig.~\ref{fig:protocol}) decays with the stiffness, as nicely predicted by the simple ABE in Eq.~\eqref{eq:sigma2-inf}.
The efficiency is thus no longer that of a Stirling process, as given in Eq.~\eqref{eq:eta-inf-eq}.

Finally, Fig.~\ref{fig:var-t-and-Teff-k} demonstrates that already partial knowledge of the variance along a full cycle of the active engine allows its mapping to the ABE.
In the figure, the mapping is achieved by adjusting minimum and maximum values of $v_a$ in Eq.~\eqref{eq:sigma2-inf} so that minimum and maximum values of $\sigma_{\infty}$ agree with those measured for $\sigma$ (illustrated by the two black dots in Fig.~\ref{fig:var-t-and-Teff-k}(a)), keeping the other parameters fixed at the values given in Sec.~\ref{sec:model}.
Using this method, the variances and effective temperatures of the simulation and the ABE are seen to agree nicely over the whole cycle (see the dashed and continuous lines in Fig.~\ref{fig:var-t-and-Teff-k}(a) and (b)).

\begin{figure}
    \centering
    \includegraphics[width=\linewidth]{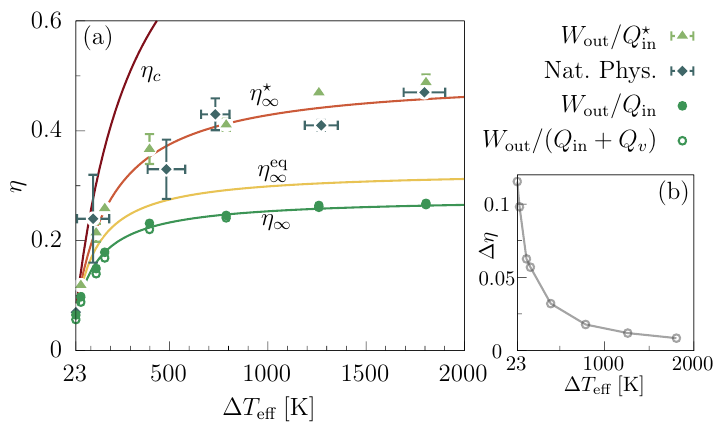}
    \caption{
    (a) Efficiencies obtained from our simulations (triangles and circles) compared to experiments~\cite{krishnamurthy2016micrometre} (diamonds) as functions of the effective temperature difference $\Delta T_{\text{eff}}$. 
    Lines depict the corresponding analytical predictions \eqref{eq:eta-inf-star} and \eqref{eq:eta-inf} from the reduced ABE model, except the dark red line (Carnot's efficiency, equivalent to Stirling's at perfect recuperation) and the yellow line (Stirling efficiency without recuperation).
    Triangles correspond to our numerical data, if employed to evaluate the (inappropriate) nominal recuperation definition of efficiency~\eqref{eq:eta-star}. 
    Full and open circles represent the most common definition of overdamped efficiency~\eqref{eq:eta} without and with considering the heat leaked through momentum degrees of freedom, respectively. 
    The relative difference $\Delta \eta$ between these two conventions is shown in (b).
    }
    \label{fig:eta-vs-Teff}
\end{figure}

Our second main result is shown in Fig.~\ref{fig:eta-vs-Teff}(a), where we present the experimentally reported efficiency from Ref.~\cite{krishnamurthy2016micrometre} (diamonds) and numerical results for the various definitions of the efficiency discussed in Sec.~\ref{sec:theory} (other symbols) together with corresponding analytical predictions from the reduced ABE model (lines).
The most common definition~\eqref{eq:eta} of efficiency for overdamped stochastic heat engines ($W_{\text{out}}/Q_{\text{in}}$, full circles), neglecting kinetic energy but accounting for the heat absorbed by the working medium during the isochoric heating branch of the Stirling cycle in Fig.~\ref{fig:protocol}, yields values slightly below the standard Stirling efficiency $\eta_{\infty}^{\text{eq}}$ in Eq.~\eqref{eq:eta-inf-eq}, whenever the bath is active. 
As explained above, this is because the actual cycle is not a perfect Stirling cycle in the $T_{\text{eff}}-k$ plane.
The numerical results for the many-body setup 
almost perfectly agree with the corresponding analytical result $\eta_\infty$ for the effective single-particle ABE in Eq.~\eqref{eq:eta-inf}.
The efficiency $W_{\text{out}}/(Q_{\text{in}} + Q_v)$ (open circles), which does additionally account for the kinetic energy contribution, is about $11\%$ smaller than $W_{\text{out}}/Q_{\text{in}}$ for the conventional ``passive'' engine ($\Delta T=\SI{23}{\kelvin}$) and quickly converges to $W_{\text{out}}/Q_{\text{in}}$ with increasing activity.
This is shown in Fig.~\ref{fig:eta-vs-Teff}(b), where we depict the relative error $\Delta \eta=1 - Q_{\text{in}}/(Q_{\text{in}}+Q_v)$ of $W_{\text{out}}/Q_{\text{in}}$. 
The heat leakage through momentum degrees of freedom can thus be safely neglected for strongly active systems. 
Moreover, $\Delta \eta=0$ when one can experimentally realize the activation at constant solvent temperature $T$, e.g., by modulating the activity via external electric fields~\cite{krishnamurthy2023overcoming-power-efficiency, rica2024heating-and-cooling}.

The efficiencies computed using the optimistic (but unrealistic) recuperation definition~\eqref{eq:eta-star} ($W_{\text{out}}/Q_{\text{in}}^\star$, triangles), nicely follow the corresponding analytical ABE prediction~\eqref{eq:eta-inf-star}, $\eta_\infty^\star$. 
The mismatch for large $\Delta T_{\text{eff}} \gtrsim \SI{1000}{\kelvin}$ is caused by the above-mentioned inaccuracy of the ABE effective temperature compared to that of our simulation, as depicted in Fig.~\ref{fig:Teff-methods}.

Interestingly, the values of the nominal efficiency at perfect recuperation, both for our simulations and the reduced analytical ABE model nicely agree with the efficiencies reported experimentally in Ref.~\cite{krishnamurthy2016micrometre}. 
\textcolor{black}{
 Our results thus strongly suggest that the authors of Ref.~\cite{krishnamurthy2016micrometre} have adopted the recuperation definition of efficiency $\eta^\star=W_{\rm out}/Q_{\rm in}^\star$ as stated in their methods section, instead of the efficiency definition $\eta=W_{\rm out}/Q_{\rm in}$ which would fit the discussion in the main body of their paper.
}
This is arguably a more straightforward and theoretically better-supported explanation of the reported unconventionally large efficiencies than those we have found in the literature \cite{krishnamurthy2016micrometre, park2022effects-of-non-markovianity, park2020active-reservoirs,park2022effects-of-non-markovianity, chang2023stochastic-heat-engines, albay2023engineered-active-noise, Kwon2024}.

\section{Conclusion and Outlook}
\label{sec:conclusion}

We performed numerical simulations of a relatively simple but physically sound many-body model of the bacterial heat engine realized experimentally in Ref.~\cite{krishnamurthy2016micrometre}.
The colloidal particle trapped by optical tweezers, which represents the working medium of the engine, was modeled as an overdamped Brownian particle in a harmonic potential.
The bacterial bath surrounding the colloid was realized as an ensemble of active Brownian particles in a Markovian solvent, interacting with the probe by a soft, short-range repulsive potential.  

The central results of our study are displayed in Figs.~\ref{fig:var-t-and-Teff-k} and~\ref{fig:eta-vs-Teff}. 
Figure~\ref{fig:var-t-and-Teff-k} contains all information about the system's average response to the imposed periodic protocol (Fig.~\ref{fig:protocol}) needed for a thermodynamic analysis of the cycle using the key concept of an effective temperature.
Importantly, the figure shows that this information is sufficiently accessible also with the limited resolution of the experimental setup of Ref.~\cite{krishnamurthy2016micrometre}. 
Figure~\ref{fig:eta-vs-Teff} illustrates how the bounds on the efficiency according to the second law limit the efficiency obtained from simulations in terms of the effective temperature.
Moreover, the puzzlingly large efficiencies reported from the experiments in Ref.~\cite{krishnamurthy2016micrometre}, which seemed in conflict with the second law of thermodynamics, are seen to nicely agree with the nominal ideal recuperation efficiency of the numerical model.
Although, in the absence of a technical implementation of such recuperation for the bacterial bath, this figure of merit has so far no practical significance, we thereby provide a simple explanation for a long-standing puzzle.
Finally, our numerical results for efficiencies and effective temperatures align well with results obtained analytically by means of the hugely simplified ABE model of the bacterial heat engine, based on a single effective active Brownian or active Ohrstein-Uhlenbeck particle.
This empirical mapping provides a nice demonstration that all systems with the same Hamiltonian and the same effective temperature yield identical average thermodynamics, including the efficiencies, for a given protocol~\cite{holubec2020active-brownian-heat-engines}.

More generally, this study shows that whenever an effective temperature $T_{\text{eff}}$ according to the second law can be defined for an engine operating with a non-equilibrium bath, it offers the most straightforward way to assess its average thermodynamic performances and to derive bounds on it. \textcolor{black}{This includes deriving optimal protocols for the output power of heat engines with active reservoirs from the known results for engines with equilibrium reservoirs~\cite{holubec2020active-brownian-heat-engines}. 
If one takes into account the housekeeping heat as in Eq.~\eqref{eq:eta_tot_full}, generalizing results on maximum efficiency~\cite{holubec2022optimal-finite-time} in the same way is not that straightforward~\cite{Fodor2024} but could be an interesting technical task for future research.} More general proposals for the derivation of bounds on the efficiency of active heat engines, such as those based on information theory~\cite{pietzonka2022second-law-for-active-heat-engines}, usually require some statistical knowledge of the system dynamics, e.g., its stationary probability distribution, which is much harder to obtain, already in the case of a single active particle~\cite{franosch2022analytic-solution} than the simple average required for the calculation of $T_{\text{eff}}$ within the present model. 

\section*{Acknowledgements}
V.H. acknowledges the support of Charles University through project PRIMUS/22/SCI/009. K.K. acknowledges financial support from the Deutsche Forschungsgemeinschaft for computational resources.

\appendix
\section{Autocorrelations of the active noise term}
\label{app:acf}
The position variance $\sigma$ of the tracer particle can in principle be computed from the noise autocorrelation function (ACF) $\langle\bm\eta(t)\cdot\bm\eta(0)\rangle$~\cite{holubec2020active-brownian-heat-engines}. 
The corresponding formula for the effective temperature is
\begin{equation}
    T_{\text{eff}} = \frac{1}{2\mu k_B} \int_0^{\infty}\langle\bm\eta(t)\cdot\bm\eta(0)\rangle e^{-\mu kt}\,\mathrm dt\,.
    \label{eq:Teff-acf}
\end{equation}
Although of theoretical interest, this Green-Kubo type relation is most likely not a practical tool for obtaining $T_{\rm eff}$ from experiments due to the required high temporal resolution. 
For this reason, instead of $\Delta t=\SI{2}{\milli\second}$, used in the rest of the paper, we utilized all simulation data points generated with timestep $\mathrm dt = \SI{20}{\micro\second}$ for the numerical evaluation of Eq.~\eqref{eq:Teff-acf}.
In our simulations, we sampled the noise ACFs at constant stiffness $k$ and temperature $T$ and checked that the results were not visibly dependent on the trap stiffness, in accord with the fact that the active bath does not feel the potential directly but only via the probe. The $k$-dependence of the effective temperature thus originates solely from the exponential term in Eq.~\eqref{eq:Teff-acf}. 

\begin{figure}[t]
    \centering
    \includegraphics[width=\linewidth]{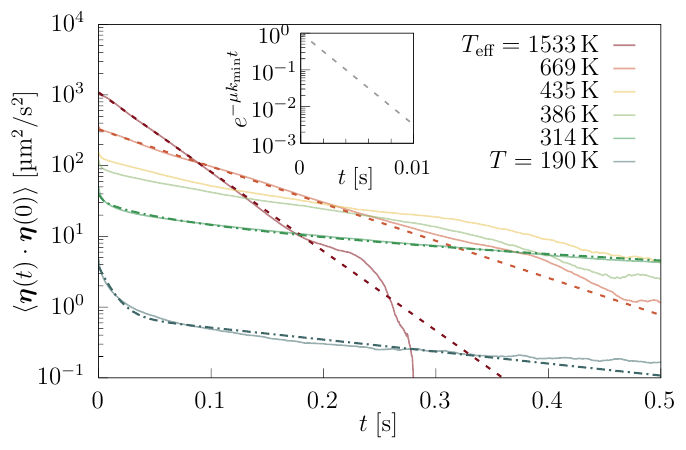}
    \caption{Autocorrelation functions of the active component $\bm\eta = \mu \bm F$ of the noise in Eq.~\eqref{eq:eom-probe}. Dashed lines represent simple exponential fits, and dash-dotted lines represent the weighted sum~\eqref{eq:sterchExp}
    of an exponential and a stretched exponential with stretching exponent $1/2$. 
    The inset shows the weighting factor to determine $T_{\text{eff}}$ via Eq.~\eqref{eq:Teff-acf}, an exponential decay with the probe's relaxation time in the harmonic potential with stiffness $k = k_{\min}$, which effectively cuts off the effect of the stretched tails on $T_{\rm eff}$.}
    \label{fig:SM-acfs}
\end{figure}

To check the influence of the background thermal noise contribution on $T_{\text{eff}}$, we evaluated the effective temperature both using the full ACF and also the ACF of only the active component of the noise, $\mu \bm F$ in Eq.~\eqref{eq:eom-probe}. 
The results for $T_{\rm eff}$ are compared in Fig.~\ref{fig:Teff-methods} of the main text. 
A few example ACFs are shown in Fig.~\ref{fig:SM-acfs}.
At high activity, they decay approximately exponentially, as indicated by the dashed lines.
The equilibrium and low-activity ACFs instead exhibit a two-step decay and are better approximated by a weighted sum of an exponential and a stretched exponential (dash-dotted lines),
\begin{equation}
    c_1 e^{-\sqrt{t / \tau_1}} + c_2 e^{-t / \tau_2}\,,
    \label{eq:sterchExp}
\end{equation}
with two time scales differing roughly by two orders of magnitude, $\tau_2/\tau_1\approx 10^2$ and $\tau_1\approx \SI{e-2}{\second}$.
The same functional form of the noise ACF has previously been observed in a similar system with a different interaction potential \cite{speck2023effective-dynamics}.

The time $1/{\mu k}$ for the probe's equilibration in the trap is orders of magnitude shorter than the decay of the $\bm\eta$-ACFs, as shown in the inset of Fig.~\ref{fig:SM-acfs} for $k=k_{\text{min}}$. 
This means that the $T_{\text{eff}}$ estimate via Eq.~\eqref{eq:Teff-acf} is dominated by the autocorrelation at very short times.
In the realm of P{\'e}clet numbers for which the exponential part of the ACF dominates one can establish a mapping to the single-particle ABE.
For $t>0$ (ignoring the $\delta$-peak at $t=0$ due to thermal noise), the ABE parameters $v_a$ and $D_r$ can be determined by fitting $C(t) = v_a^2e^{-D_rt}$ to the steady-state ACFs at $(k_{\max}, v_{\min})$ and $(k_{\min}, v_{\max})$, corresponding to the black points in Fig.~\ref{fig:var-t-and-Teff-k}(a).
This way, an effective time-dependence in $D_r$ is introduced, similar to those for the temperature and the P\'{e}clet number in Fig.~\ref{fig:protocol}. 

\section{Stochastic efficiency}
\begin{figure}
    \centering
    \includegraphics[width=\linewidth]{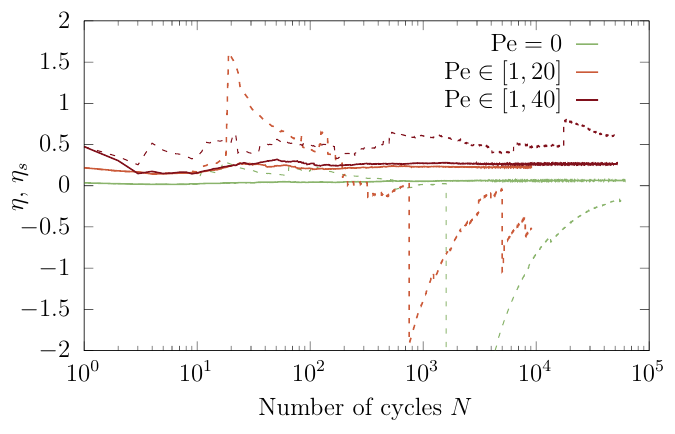}
    \caption{
    Thermodynamic efficiency $\eta=\langle -w\rangle_N/\langle q_{\text{in}}\rangle_N$ (solid lines) and the (ill-defined) average stochastic efficiency $\eta_s=\langle -w / q_{\text{in}}\rangle_N$ (dashed lines) as functions of the number $N$ of cycles used to evaluate the averages, for the passive and two active realizations of the numerical model.
    While the thermodynamic efficiency, defined as the ratio of averages, converges relatively fast, the average of the stochastic ratio does not exist. 
    }
    \label{fig:SM-eta-average}
\end{figure}

The stochastic work and heat currents during one cycle are defined analogously to their averages in Sec.~\ref{sec:theory}, just the averaging over the noise is not imposed.
The definitions of stochastic work $w_{\text{out}}$, delivered by the system, configurational input heat $q_{\text{in}}$ (without the kinetic energy contribution), and ideal-recuperation input heat $q_{\rm in}^\star$ read
\begin{eqnarray}
w_{\text{out}}(t) &=& -\frac{1}{2}\int_0^t\dot{k}(t') \bm r^2(t') \,\mathrm dt'\,,
\label{eq:q_in-star11}\\
q_{\text{in}}(t) &=& \int_0^t k(t') \bm r(t')\cdot\dot{\bm r}(t') \Theta(\dot{\sigma})\,\mathrm dt'\,,
    \label{eq:q_in}\\
q_{\text{in}}^\star(t) &=& \int_0^t k(t') \bm r(t')\cdot\dot{\bm r}(t') \delta_{T_{\max}, T(t)}\,\mathrm dt'\,.
    \label{eq:q_in-star}		
\end{eqnarray}    
In Fig.~\ref{fig:SM-eta-average}, time averages $\eta_s= \langle w_{\rm out}/q_{\rm in} \rangle_N$ of the stochastic efficiency are shown as functions of the sample size $N$ \textcolor{black}{(i.e., $t=N t_p$ in \eqref{eq:q_in-star11}--\eqref{eq:q_in-star})}, compared to the proper thermodynamic efficiency $\eta = \langle w_{\rm out}\rangle_N / \langle q_{\rm in}\rangle_N$.
While $\eta$ converges rapidly after about $N\approx 10^2$ cycles, $\eta_s$ never converges, due to the large fluctuations of $q_{\text{in}}$ in the denominator, which can introduce huge jumps at arbitrary $N$.

\label{app:eta}

\section{ABP and AOUP models}
\label{apx:ABPAOUP}
The dynamical equations for an AOUP in a harmonic potential are
    \begin{align}
    \begin{split}
        \dot{\bm r} &= -\mu k{\bm r} + {\bm v}  + \sqrt{2D_t}\bm\xi\,,\\
        \tau\dot{{\bm v} } &= -{\bm v} + \sqrt{2D_a}\bm\nu\,.\\
    \end{split}
\end{align}
Both of these equations are linear in the dependent variables ${\bm r}$ (position of AOUP) and ${\bm v}$ (active self-propulsion of AOUP), and thus the solution $({\bm r},{\bm v})$ is a linear functional of the Gaussian noise variables $\bm{\xi}$ and $\bm{\nu}$. Since any linear combination of Gaussian random variables is Gaussian unless an initial condition destroys Gaussianity, and we study the AOUP only in the steady state, where the initial perturbations from Gaussianity are forgotten, the AOUP model is, for our purposes, Gaussian.

The dynamical equations for an ABP in a harmonic trap in two dimensions read
\begin{align}
    \begin{split}
        \dot{\bm r} &= -\mu k{\bm r} + v_a\bm n + \sqrt{2D_t}\bm \xi\,,\\
        \dot{\theta} &= \sqrt{2D_r}\nu\,,
    \end{split}
\end{align}
where $\bm n = (\cos \theta, \sin \theta)$ is the orientation vector of the ABP. In this case, the dynamical equations are nonlinear in the orientation angle $\theta$, which is the source of non-Gaussianity of the ABP model even in the steady state, whenever $v_a>0$ and $D_r < \infty$. 

The average thermodynamics of the bacterial heat engine in terms of heat and work is solely described by the position variance $\sigma = \langle {\bm r}^2 \rangle$. 
In both the ABP and AOUP models the position obeys the dynamical equation $\dot{\bm r} = -\mu k {\bm r} + \bm \eta$, where $\bm \eta$ is an exponentially correlated noise with the exponential autocorrelation functions $\langle\eta_i(t) \eta_j(t')\rangle = \delta_{ij}v_a^2 e^{-D_r|t-t'|} + 2D_t\delta_{ij}\delta(t-t')$ for the ABP and $\langle\eta_i(t) \eta_j(t')\rangle = \delta_{ij}2D_a e^{-|t-t'|/\tau}/\tau + 2D_t\delta_{ij}\delta(t-t')$ for the AOUP. 
For the ABP, the noise $\bm \eta$ is non-Gaussian, and for the AOUP, it is Gaussian.
However, since the position variance, $\langle {\bm r}^2 \rangle$, is determined just by the noise autocorrelation function according to Eq.~\eqref{eq:Teff-acf} and is independent of its higher moments, both the AOUP and ABP models yield the same average thermodynamic performance for $\tau = 1/D_r$ and $v_a^2=2 D_a$.

The difference between the two models becomes visible in the fluctuations of position, work, and heat, as demonstrated in Ref.~\cite{partII}.

\bibliographystyle{apsrev4-2}
\bibliography{refs}
\end{document}